\documentclass[prb,a4paper,twocolumn,floatfix,showpacs,showkeys,amsmath,amssymb,nobibnotes,unsortedaddress,superscriptaddress]{revtex4-1}
\usepackage{graphicx}
\usepackage[utf8]{inputenc}

\begin{document}

\title{Homocoupling defects in a conjugated polymer limit exciton diffusion}

\author{Martin Streiter}
\affiliation{Institut für Physik, Technische Universität Chemnitz, 09126 Chemnitz, Germany}
\author{Daniel Beer}
\affiliation{Institut für Physik, Technische Universität Chemnitz, 09126 Chemnitz, Germany}
\author{Fabian Meier}
\affiliation{Institut für Physik, Technische Universität Chemnitz, 09126 Chemnitz, Germany}
\author{Clemens Göhler}
\affiliation{Institut für Physik, Technische Universität Chemnitz, 09126 Chemnitz, Germany}
\author{Caroline Lienert}
\affiliation{Institut für Makromolekulare Chemie, Universität Freiburg, 79104 Freiburg, Germany}
\affiliation{Institut für Chemie, Technische Universität Chemnitz, 09126 Chemnitz, Germany}
\author{Florian Lombeck}
\affiliation{Institut für Makromolekulare Chemie, Universität Freiburg, 79104 Freiburg, Germany}
\affiliation{Cavendish Laboratory, University of Cambridge, UK}
\author{Michael Sommer}
\affiliation{Institut für Makromolekulare Chemie, Universität Freiburg, 79104 Freiburg, Germany}
\affiliation{Institut für Chemie, Technische Universität Chemnitz, 09126 Chemnitz, Germany}
\author{Carsten Deibel}
\affiliation{Institut für Physik, Technische Universität Chemnitz, 09126 Chemnitz, Germany}
\email{deibel@physik.tu-chemnitz.de}

\begin{abstract}
Copolymers  such as PCDTBT (poly(N-9'-heptadecanyl-2,7-carbazole-\textit{alt}-5,5-(4',7'-di-2-thienyl-2',1',3'-benzothiadiazole))) are commonly employed as donor material in bulk heterojunction solar cells. Recently, chemical defects such as homocouplings have been shown to form at the material synthesis stage, strongly reducing the short circuit current in organic photovoltaics. Here we show that both, low molecular weight and homocoupling defects reduce the short circuit current of solar cells because of limited exciton diffusion. We propose a model that unites and explains the influence of both chemical parameters with the distribution of conjugation lengths. The  connection between limited exciton diffusion and short circuit current is revealed by kinetic Monte Carlo simulation of bulk heterojunctions. Our findings are likely applicable for copolymers in general.
\end{abstract}

\keywords{exciton diffusion; organic semiconductors; photovoltaics; homocoupling; PCDTBT}

\maketitle

\section{Introduction}

Exciton diffusion is crucial for organic photovoltaic devices (OPV) as singlet excitons are the primary photoexcitations in organic semiconductors.\cite{serdar1998primary} Copolymers with  donor-acceptor structure are a promising material class in organic photovoltaics. Their optoelectronic properties can be controlled by pairing different donor and acceptor subunits. A strict alternation of donor and acceptor subunits is commonly assumed. This  assumption has been challenged by recent literature.\cite{Hendriks2014,Lombeck2016} The copolymer PCDTBT (poly(N-9'-heptadecanyl-2,7-carbazole-\textit{alt}-5,5-(4',7'-di-2-thienyl-2',1',3'-benzothiadiazole))) consists of alternating  carbazole (cbz) and thiophene-benzothiadiazole-thiophene (TBT) groups. The connection between its molecular and optoelectronic properties are well studied.\cite{Gieseking2012,biskup2015ordering,Lombeck2016,matt2018tbt,banerji2010exciton,Banerji2012} Lombeck \textit{et al}.\ showed that most synthesis protocols for PCDTBT lead to significant concentrations of carbazole homocoupling.\cite{Lombeck2016} Hendriks \textit{et al}.\cite{Hendriks2014}\ and Lombeck \textit{et al}.\cite{Lombeck2016}\ have demonstrated lower photocurrents  in solar cells with diketopyrrolopyrrole-based copolymers and PCDTBT, respectively. Both studies suggested that reduced exciton dissociation caused by localized levels of lowest unoccupied molecular orbitals (LUMO) may explain the decrease of photocurrent. Here, for the first time, we report on the a  dependence of exciton diffusion on homocoupling density. Exemplified by a series of PCDTBT samples with varying carbazole homocoupling concentration and molecular weight, we show that exciton diffusion is limited when the effective conjugation length of the polymer chain is decreased. Our findings further  emphasize the importance of controlled synthesis accompanied by careful analysis.

In PCDTBT, both the cbz and TBT groups can form homocoupling bonds, but  TBT homocoupling is negligible.\cite{Lombeck2016} We therefore refer to carbazole homocoupling (cbz-cbz) simply as homocoupling (hc) in the following. The hc concentration was  determined  experimentally by NMR spectroscopy (see Methods section and Supporting Information) and refers to the total number of homocoupling bonds in the chain divided by all bonds (alternating bonds $+$ homocoupling bonds). 

A variety of techniques for measuring exciton diffusion is applied in the literature.\cite{Lin2014,Mikhnenko2015} We used the bulk quenching method to determine the  exciton diffusion coefficient and length at room temperature. This technique compares photoluminescence (PL) decay of thin films with different concentrations of embedded quencher molecules.\cite{Mikhnenko2012} As quenchers, we used  PC$_{61}$BM (phenyl-C61-butyric acid methyl ester) assuming a quenching efficiency of unity and a  quenching radius $r\approx1\,\mathrm{nm}$.\cite{Mikhnenko2012,Lin2015} We  extracted the corresponding PL lifetime $\tau$  with $PL(t)\propto \exp(-t/\tau)$. Measuring $\tau$ as a function of the  quencher  concentration $c$ yields the exciton diffusion coefficient $D$ with the  Stern--Volmer equation,\cite{powell1975,Lin2014}
\begin{equation}
    \frac{1}{\tau(c)}=\frac{1}{\tau_0} + 4\pi r D c.
    \label{sv}
\end{equation}
The three-dimensional ($Z=3$) diffusion length\cite{powell1975} $L_{\mathrm{D}}$ results from the radiative lifetime $\tau_0$ of the pristine material  and $D$. Note that the  factor 2  is sometimes omitted and that often the one-dimensional case is considered in the literature.
\begin{equation}
    L_{\mathrm{D}} = \sqrt{2ZD\tau_0}
     \label{length}
\end{equation}

\section{Results}

\begin{figure}[ht]
\includegraphics*[scale=0.6]{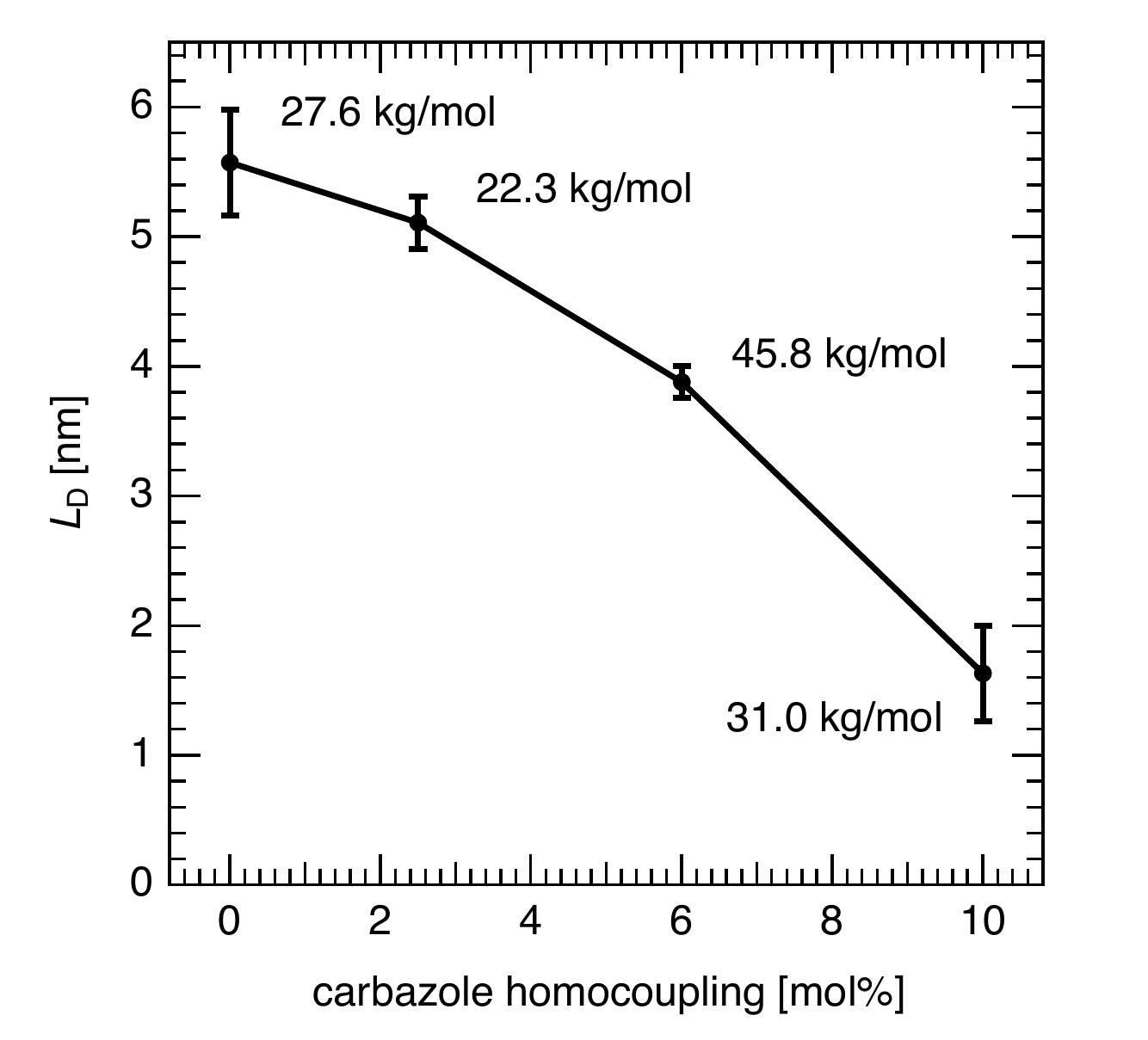}
\caption{Diffusion length $L_{\mathrm{D}}$ as a function of carbazole homocoupling concentration in PCDTBT films.}
\label{DL_hk}
\end{figure}

We first compared exciton diffusion in  films  between $0\,\%$ and $10\,\%$ hc samples with comparable number average molecular weight ($M_{\mathrm{n,SEC}} = 27.6\,\mathrm{kg\,mol^{-1}}$ and $31.0\,\mathrm{kg\,mol^{-1}}$, respectively). PL decay curves and Stern--Volmer plots are shown in the Supporting Information. Fitting the PL lifetimes of films with  different PCBM concentrations with the Stern--Volmer equation (\ref{sv}) yielded an exciton diffusion coefficient  $D=0.63\cdot10^{-4}\,\mathrm{cm^2 s^{-1}}$ for $0\,\%$  hc and $D=0.06\cdot10^{-4}\,\mathrm{cm^2 s^{-1}}$ for $10\,\%$  hc.  Ward \textit{et al}.\ found  $D = (1.1\pm0.5) \cdot10^{-4}\,\mathrm{cm^2~s^{-1}}$ in PCDTBT.\cite{Ward2012} Note that homocoupling concentration and molecular weight were not reported. We additionally measured diffusion in a commercial PCDTBT batch  with $2.5\,\%$ hc, $M_{\mathrm{n,SEC}} = 22.3\,\mathrm{kg\,mol^{-1}}$ and a batch with $6\,\%$ hc, $M_{\mathrm{n,SEC}} = 45.8\,\mathrm{kg\,mol^{-1}}$. The molecular weight   of both batches deviated from the  $0\,\%$ hc batch by about  $-20\,\%$ and $+50\,\%$, respectively. Nevertheless, our results for $D$ fitted the trend that the diffusion length decreases for higher homocoupling concentration, as shown in figure \ref{DL_hk}. 

\begin{figure}[ht]
\includegraphics*[scale=0.6]{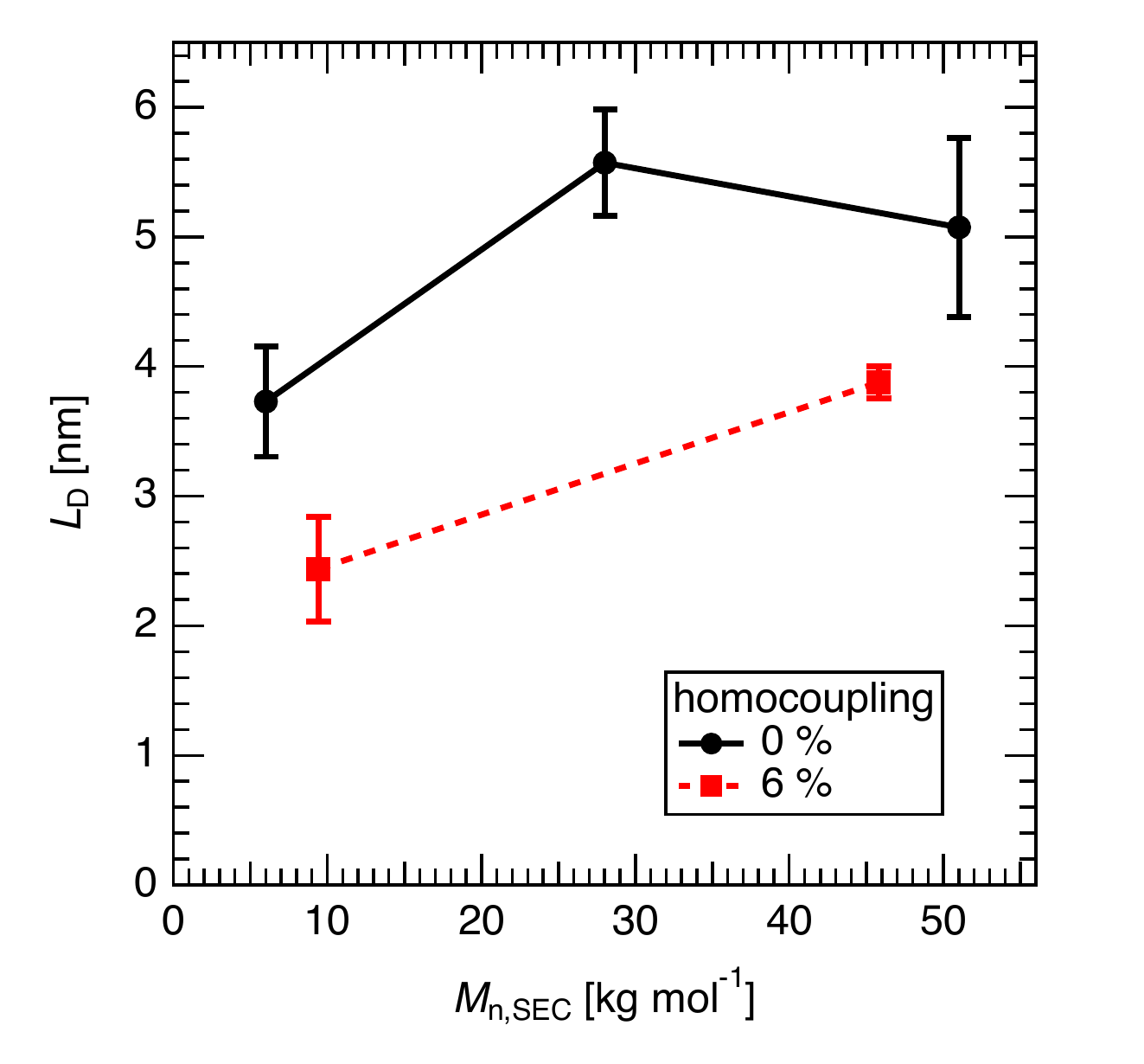}
\caption{Influence of molecular weight on  diffusion length $L_{\mathrm{D}}$   in PCDTBT films with $0\,\%$ and $6\,\%$ homocoupling.}
\label{DL_Mn}
\end{figure}

To assess the possible influence of the mentioned molecular weight deviations, we measured $D$ for three $0\,\%$ hc  and two $6\,\%$ hc batches with varying molecular weight. Figure \ref{DL_Mn} shows that the lowest molecular weight resulted in the shortest diffusion length. At $0\,\%$ hc, we found a maximum at  $27.6\,\mathrm{kg\,mol^{-1}}$. It is not clear if such maximum exists in the case of $6\,\%$ hc. However, we observed  that low molecular weights decrease $L_{\mathrm{D}}$ with and without homocoupling in the polymer chain. With these findings, we can  reevaluate figure \ref{DL_hk}.  The $L_{\mathrm{D}}$ of the   $2.5\,\%$ hc batch is underestimated because of the $20\,\%$ lower $M_{\mathrm{n,SEC}}$. In contrast,    $L_{\mathrm{D}}$  of the $6\,\%$ hc  batch is overestimated because of the $50\,\%$ higher $M_{\mathrm{n,SEC}}$. In summary,  we find evidence for  monotonously decreasing exciton diffusion length with increasing carbazole homocoupling concentration in PCDTBT at a given molecular weight.

\section{Discussion}
We now discuss  the interaction between excitons and homocoupling defects by evaluating quenching, traps and the distribution of homocoupling defect sites in the chain.

The bulk quenching method compares  PL lifetimes of a pristine and quenched film, as described above.  We  hypothetically consider carbazole homocoupling defects as perfect quenchers with the same properties as PCBM molecules (quenching radius $\approx1\,\mathrm{nm}$, homogeneously distributed in the film). From the PCDTBT film density\cite{mateker2015}  of $\rho = 1.16\,\mathrm{g\,cm^{-3}}$ and $M_{\mathrm{n,SEC}} = 31.0\,\mathrm{kg\,mol^{-1}}$ follows a homocoupling defect concentration of $c_{\mathrm{hc}}\approx10^{20}\,\mathrm{cm^{-3}}$ in the  $10\,\%$  hc film. Adding this amount of $10^{20}\,\mathrm{cm^{-3}}$ quenchers to a $0\,\%$  hc film   would decrease its PL lifetime from $\tau_0 = 828\,\mathrm{ps}$ to 0.1\,ps according to the Stern--Volmer equation (\ref{sv}). However, the  lifetime of the $10\,\%$  hc film showed only a small reduction to  $\tau_0 = 699\,\mathrm{ps}$. We therefore conclude that the interaction between homocoupling defects and excitons  cannot be described as quenching. 

Several studies have shown that trap states decrease the diffusion length in organic semiconductor films.\cite{Mikhnenko2014, Rorich2017, Anthanasopoulos2008} PCDTBT films have low structural order and are amorphous.\cite{Beiley2011, Cho2010} Exciton diffusion is  isotropic in PCDTBT.\cite{Ward2012} We assume that exciton diffusion can be described as interchain hopping with only weak contribution of intrachain processes.\cite{schwartz2001,Schwartz2003,Markov2006, Hennebicq2005}  We now discuss if homocoupling defects act as exciton traps. Lombeck \textit{et al}.\ found that carbazole homocoupling defects cause a localization of the LUMO on the neighboring TBT groups. The highest occupied molecular orbital (HOMO) energy thereby decreases by $0.02\,\mathrm{eV}$ and the LUMO by  $0.05\,\mathrm{eV}$.\cite{Lombeck2016} If we assume homocoupling defect sites as randomly distributed trap sites in the film, we can relate the homocoupling concentration to the number of traps in the film, $\mathrm{hc} \propto c_{\mathrm{trap}}$. Athanasopoulos \textit{et al}.\ \cite{Anthanasopoulos2008} simulated the influence of a trap concentration $c_{\mathrm{trap}}$ on the diffusion length and found it to be in agreement with analytic predictions by Montroll and Weiss\cite{montroll1965} with  $L_{\mathrm{D}} \propto c_{\mathrm{trap}}^{-1/2}$. Mikhnenko \textit{et al}.\ experimentally found  $L_{\mathrm{D}} \propto c_{\mathrm{trap}}^{-1/3}$ by comparing the calculated intrinsic $c_{\mathrm{trap}}$ and its influence on diffusion for different materials.\cite{Mikhnenko2014} All studies find a strong dependence of $L_{\mathrm{D}}$ on traps, scaling with an inverse power law. However, our result (figure \ref{DL_hk}) shows only weak or no dependence of exciton diffusion length on homocoupling in the regime between $0-2.5\,\%$ hc as discussed above. Thus, we believe that considering homocoupling defects as traps is insufficient in describing the limitation of exciton diffusion. If homocoupling defects acted as traps for excitons, diffusion characteristics would be  less dependent on molecular weight and strongly correlated with  $c_{\mathrm{trap}} \propto \mathrm{hc}$ between  $0-2.5\,\%$ hc. 

We believe that homocoupling defects alter the conjugation of the electronic system along the polymer backbone. Exciton delocalization and  conjugation length are not equal.\cite{Kohler2012} However, the maximum  space available for excitons to  delocalize is confined by the effective conjugation length of the polymer.\cite{Rauscher1990,Kohler2015}   This effective conjugation length can be limited by conjugation breaks introduced by spatial disorder (e.g. chain defects, distortions).\cite{Rauscher1990,Ruseckas2005} 

\begin{figure}[ht]
\includegraphics*[scale=0.6]{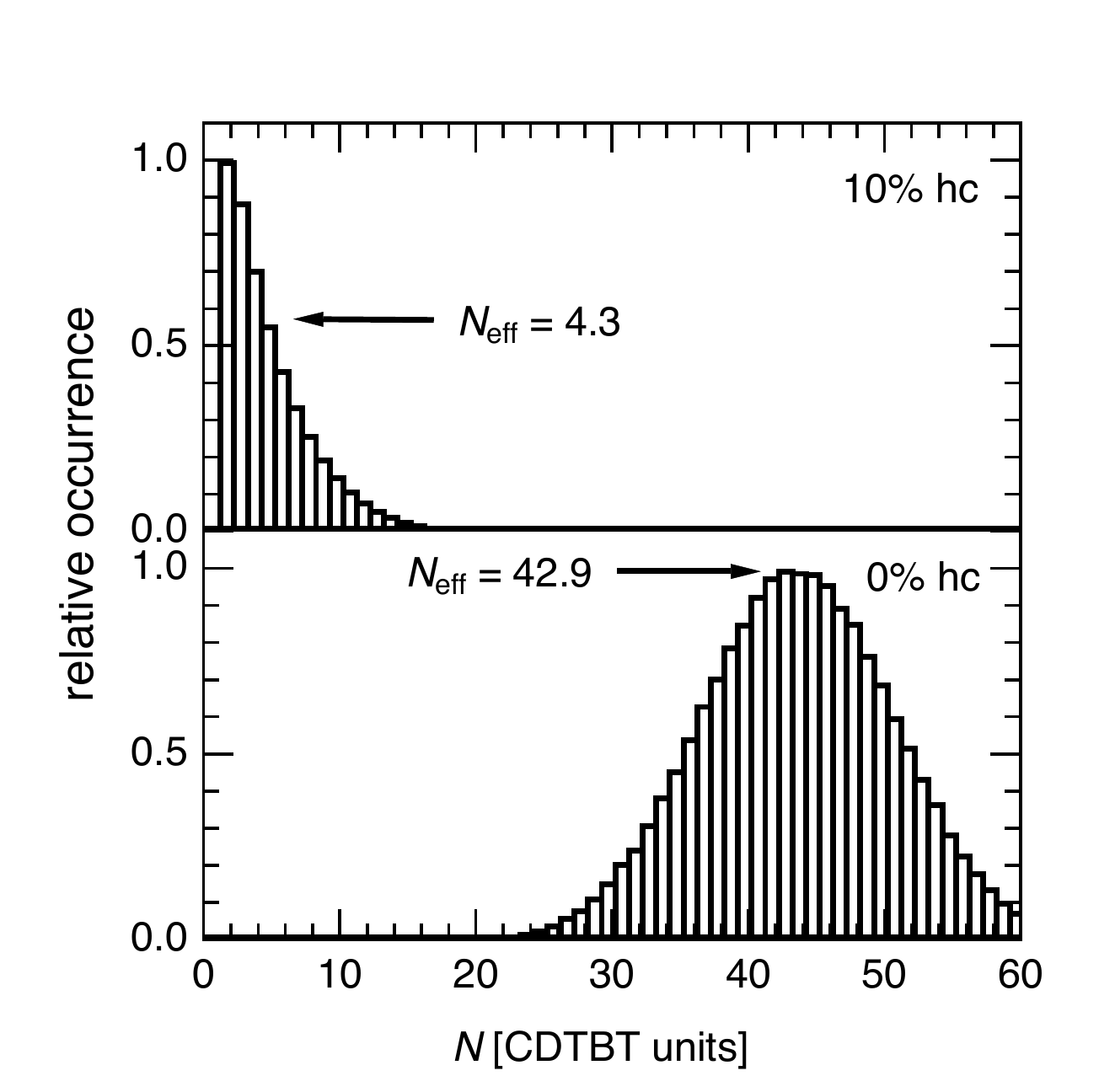}
\caption{Modelled distribution of chromophores defined as number of alternating $\mathrm{(cbz-TBT)}_N$ segments without homocoupling in the polymer chain for two different hc concentrations. }
\label{histo}
\end{figure}

Carbazole homocoupling defects lead to a localization of the LUMO as reported by Lombeck \textit{et al.}\cite{Lombeck2016} From this finding we assume that homocoupling sites act as conjugation breaks. We evaluated the resulting distribution of conjugated polymer segments (chromophores in the following) by simulating sequences of cbz and TBT in straight polymer chains. We considered carbazole homocoupling as randomly distributed occurrences of (cbz-cbz-TBT) segments within alternating (cbz-TBT)$_n$ polymer chains with Gaussian distributed molecular weight. We then evaluated the chromophore lengths by counting strictly alternating (cbz-TBT) sequences along each chain, setting the counter to zero whenever homocoupling occurred.  This estimation  assumes ideal polymer chains without spatial disorder and therefore represents the upper limit of conjugation length. Figure \ref{histo} shows modelled chromophore distributions of films with 0\,\% and 10\,\% hc at $M_\mathrm{n} = (30\pm5)\,\mathrm{kg\,mol^{-1}}$.  Homocoupling alters the Gaussian distribution of defect-free polymer chains into an exponential distribution. As the Gaussian distribution neglects spatial disorder, we expect the real distribution to include shorter chromophores as well. However, at 10\,\% hc  most chromophores are shorter than four CDTBT repeat units. Conjugation breaks caused by spatial disorder are presumably negligible in a distribution that mostly consists of few chromophores. We assigned the arithmetic mean of the modelled distribution to each PCDTBT batch. Note that the arithmetic mean of two different molecular weights with equal homocoupling concentration is smaller at lower molecular weights while the harmonic mean is equal. We chose the arithmetic mean as a measure for the average chromophore length $N_{\mathrm{eff}}$ because of the dependency of $L_{\mathrm{D}}$ on the number average molecular weight (which is an arithmetic mean as well).

Banerji \textit{et al}.\ showed that in PCDTBT, the HOMO increases by  $0.15\,\mathrm{eV}$ and  the LUMO decreases by  $0.08\,\mathrm{eV}$ when  the conjugation length is increased from one to four CDTBT units.\cite{Banerji2012} We should therefore expect different chromophore energies depending on $N_{\mathrm{eff}}$ which is determined by homocoupling and molecular weight. To probe the average chromophore energy, we measured steady state photoluminescence in solution. Measuring in solution avoided static film disorder and reabsorption effects. 

\begin{figure}[ht]
\includegraphics*[scale=0.6]{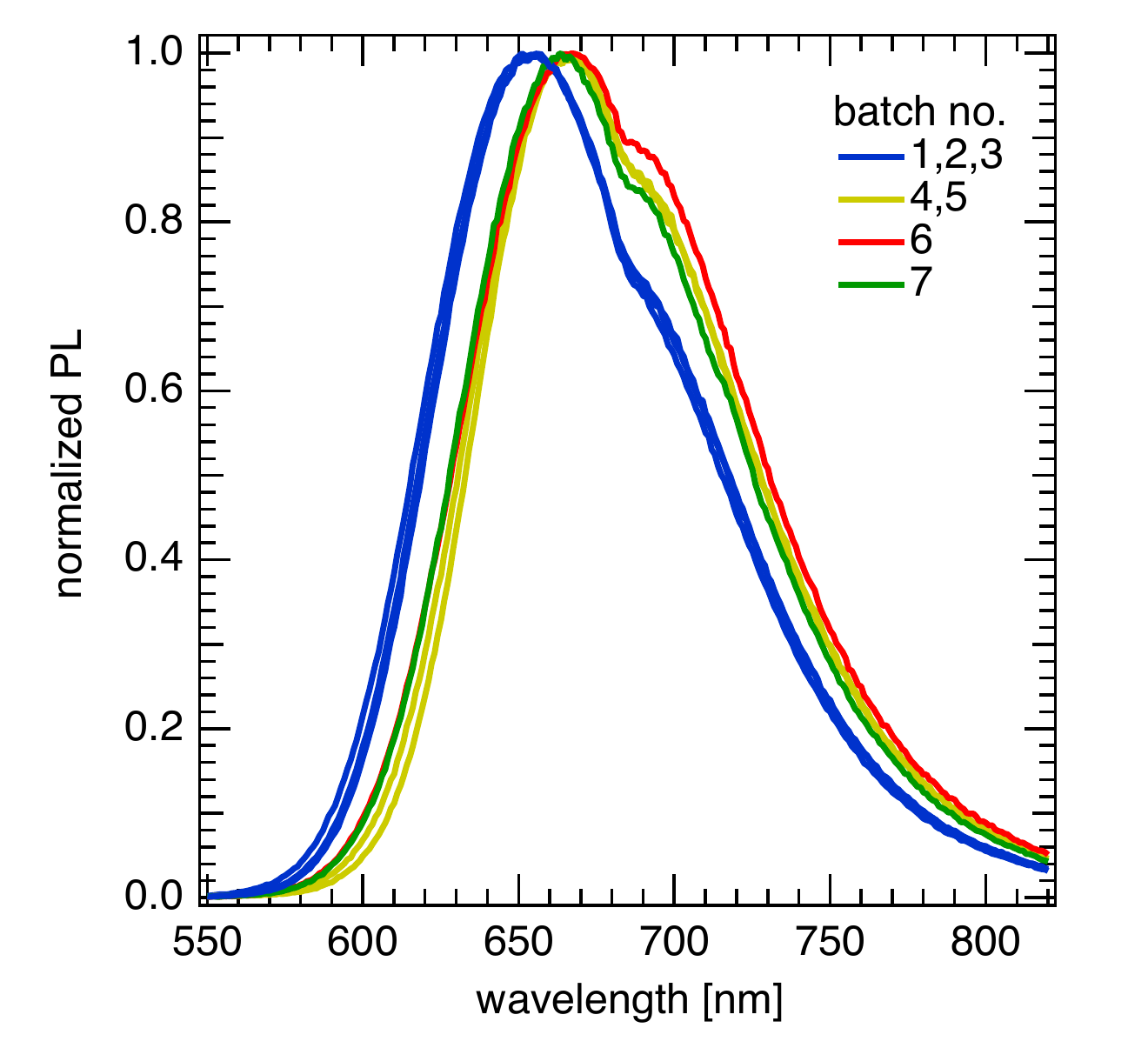}
\caption{Steady state photoluminescence of all PCDTBT batches in 1,2-dichlorobenzene solution. All spectra are normalized. Similar spectra are plotted in the same color for better readability.}
\label{PL}
\end{figure}

The PL spectra in figure \ref{PL} show a blue shift of PCDTBT batches with high homocoupling or low molecular weight (batch 1: 0\%\,hc, $6.2\,\mathrm{kg\,mol^{-1}}$, 2: 6\%\,hc, $9.4\,\mathrm{kg\,mol^{-1}}$, 3: 10\%\,hc, $31.0\,\mathrm{kg\,mol^{-1}}$). These batches have low $N_{\mathrm{eff}}$ values of 8.8, 5.3 and 4.3, respectively. Batch 7 (6\%\,hc, $45.8\,\mathrm{kg\,mol^{-1}}$) with $N_{\mathrm{eff}} = 7.2$ only shows a minor blue shift although $N_{\mathrm{eff}}$ is smaller than the 0\%\,hc, $6.2\,\mathrm{kg\,mol^{-1}}$ batch. This might be explained by the different distributions of chromophores which is not represented by $N_{\mathrm{eff}}$. Our simulation showed that this batch incorporates longer chromophores which are not existing in batch 1 and 2 because of the low molecular weight and batch 3 because of the high homocoupling concentration. Batches 4,5,6 share similar PL spectra because their large $N_{\mathrm{eff}}$ and broad  distributions of conjugation lengths lead to  similar chromophore energies. The commercial sample (batch 6) shows a more pronounced low-energy shoulder, indicating chemical deviations from our PCDTBT batches.

We now discuss the influence of  chromophore distribution on exciton diffusion. The hopping of excitons between localized sites can be described as energy downhill migration  from higher to lower localized states.\cite{Rauscher1990,mollay1994} The  distribution of these localized states is influenced by the local energetic environment, trap states and variability of conjugation length.\cite{Menke2014} In a previous work, we  showed  that even identical chromophores can have different absorption and emission energies due to the energetic disorder of the local  environment in a polymer film.\cite{streiter2016} However, we now assume similar spatial disorder, energetic disorder and trap states in the films and only  discuss the influence of chromophore  distribution. As shown in figure \ref{histo}, 10\,\% hc leads to an exponential distribution of conjugation lengths where most polymer sequences are cbz-cbz-TBT.

The influence of conjugation length on exciton diffusion was studied by Hennebicq \textit{et al}.\ \cite{Hennebicq2005} They found increased site-to-site coupling between shorter chromophores. This may explain the decrease of $\tau_0$ in films with homocoupling and short chain lengths (see Supporting Information). However, such increased coupling mostly affects chromophores with similar length and energy.\cite{Hennebicq2005} We believe that the one order of magnitude decrease of the diffusion coefficient results from downhill migration within the exponential distribution of chromophores in the 10\,\% homocoupling film. Excitons may  only hop between  few sites until they reach a longer chromophore with lower energy. These low energy chromophores can  act as a trap due to the large energy differences to surrounding shorter chromophores.\cite{Rauscher1990,Menke2014} 

In contrast, films with long chain lengths and low homocoupling concentration have broad distributions of longer chromophores. Energy differences between these longer chromophores are smaller than in films with a steep exponential distribution as shown in figure \ref{histo}. It is less likely for excitons to be trapped on an individual chromophore in a broad distribution of conjugation lengths with smaller energy differences.  Note that our simulation and calculation of conjugation lengths considers straight polymer chains without spatial disorder. In the film, we expect spatial disorder to additionally limit the conjugation length.\cite{Rauscher1990,Ruseckas2005} This explains why batches with low homocoupling and medium or high molecular weight have different values of $N_{\mathrm{eff}}$ but similar exciton diffusion lengths and PL lifetimes (see Supporting Information). The average conjugation length in these batches is $N_{\mathrm{eff}} > 10$. In this range of effective conjugation lengths, we can consider spatial disorder to limit the conjugation length, as reported in the literature.\cite{Kohler2012} To summarize, decrease of exciton diffusion occurs when the chromophore length is more limited by homocoupling or low molecular weight than by spatial disorder.   Exciton diffusion coefficient, diffusion length, PL lifetime and PL energy changed below an average chromophore length of $N_{\mathrm{eff}}=10$ CDTBT units. We explained these changes with different conjugation length distributions and chromophore energies within the polymer film. 

Another relevant aspect arises from this model. When comparing two polymer chains of equal length with and  without homocoupling, the number of chromophores per chain is different. This also applies to defect-free polymer chains with different molecular weight when considering an identical mass of these two polymers. We can estimate  the number of chromophores per film volume as  chromophore density $C_{\mathrm{cr}}$  with our simulated $N_{\mathrm{eff}}$ and the monomer molecular weight $M_{\mathrm{CDTBT}}=0.7\,\mathrm{kg\,mol^{-1}}$:

\begin{equation}
    C_{\mathrm{cr}} \approx \frac{\rho_{\mathrm{PCDTBT}}}{N_{\mathrm{eff}} \cdot M_{\mathrm{CDTBT}}}.
    \label{CR_rho}
\end{equation}

It is also possible to estimate $C_{\mathrm{cr}}$ without simulated $N_{\mathrm{eff}}$ values using only  
the  homocoupling concentration $hc$, the molecular weight of the monomer $M_{\mathrm{CDTBT}}$ and polymer batch $M_{\mathrm{n}}$:

\begin{equation}
   C_{\mathrm{cr}} \approx \frac{\rho_{\mathrm{PCDTBT}}}{ M_{\mathrm{n}}}  \left(\frac{2\textrm{hc}\cdot M_{\mathrm{n}}}{M_{\mathrm{CDTBT}}}+1\right)
     \label{meff}
\end{equation}

Note that the definition of hc may vary in the literature. Further explanations of hc and derivation of equation (\ref{meff}) are given in the Supporting Information. 

Figure \ref{ldens} shows a linear relation between diffusion length and calculated chromophore densities. We assign this trend to the inverse dependence of $C_{\mathrm{cr}}$ on $N_{\mathrm{eff}}$ (equation (\ref{CR_rho})). Larger effective conjugation lengths lead to small deviations of $C_{\mathrm{cr}}$. This can explain why two different 0\,\% hc films (27.6 and $50.7\,\mathrm{kg\,mol^{-1}}$) as well as the 2.5\,\% hc film ($22.3\,\mathrm{kg\,mol^{-1}}$) showed comparable exciton diffusion lengths. Although 2.5\,\% hc is a significant concentration of defects, $C_{\mathrm{cr}}$ is similar to films with  0\,\% hc.  The model also clarifies the non-intuitive finding that  $L_{\mathrm{D}}$ is similar in a  defect-free film with short chains (0\,\% hc, $6.2\,\mathrm{kg\,mol^{-1}}$) and a high homocoupling film with long chains (6\,\% hc, $45.8\,\mathrm{kg\,mol^{-1}}$). Both polymers share a similar value of $C_{\mathrm{cr}}$. However, as shown in the PL data of figure \ref{PL}, differences still can occur because of the  different conjugation length distributions. Batches with long chains and hc defects can include larger chromophores than defect-free short chains while having a similar chromophore density.

\section{Implications for OPV}
\begin{figure}[ht]
\includegraphics*[scale=0.7]{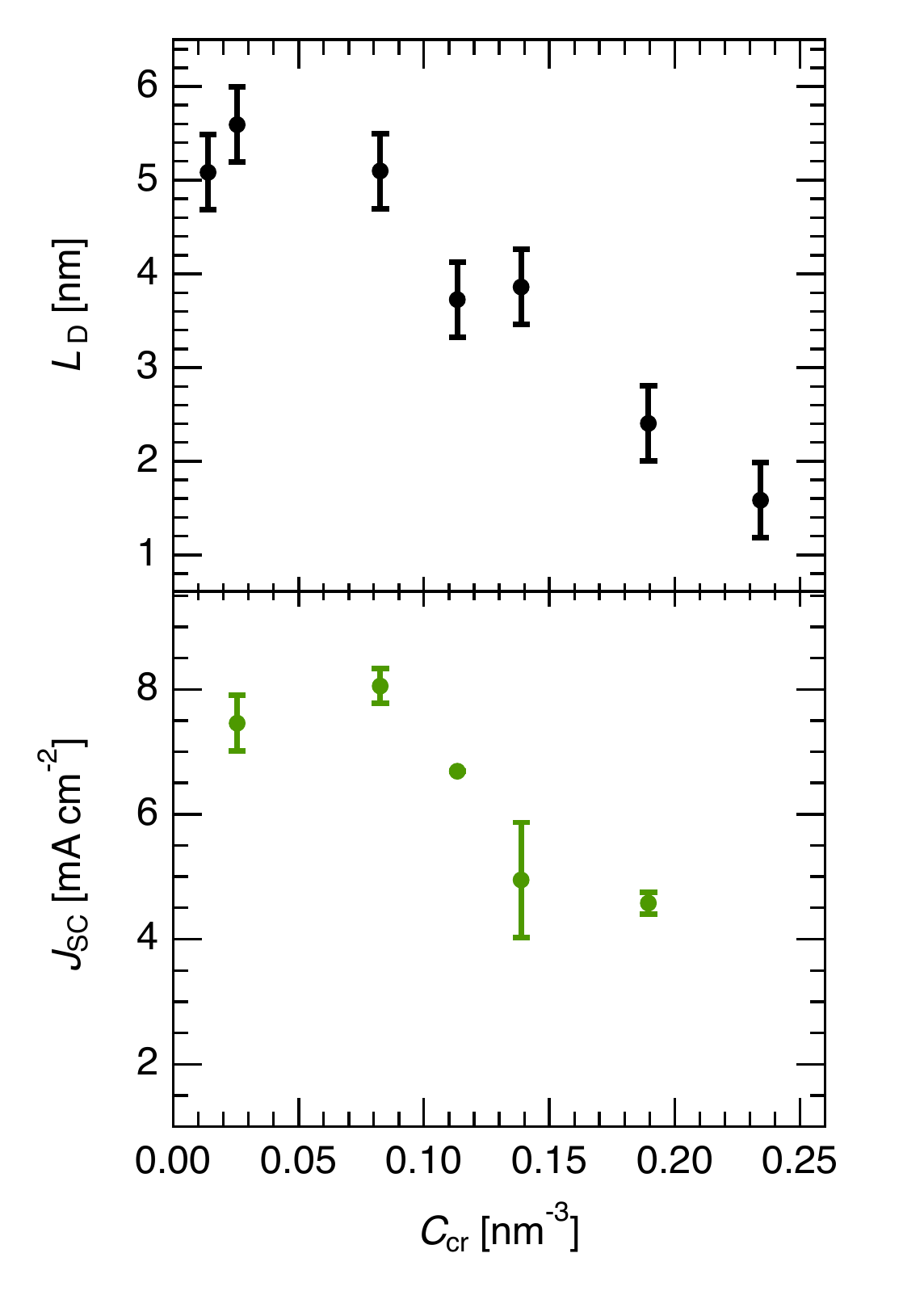}
\caption{$L_{\mathrm{D}}$ and $J_{\mathrm{sc}}$ of  PCDTBT batches with different homocoupling and molecular weight. Each polymer is assigned a chromophore density depending on homocoupling and molecular weight derived from simulated polymer chains.}
\label{ldens}
\end{figure}

\begin{figure}[ht]
\includegraphics*[scale=0.6]{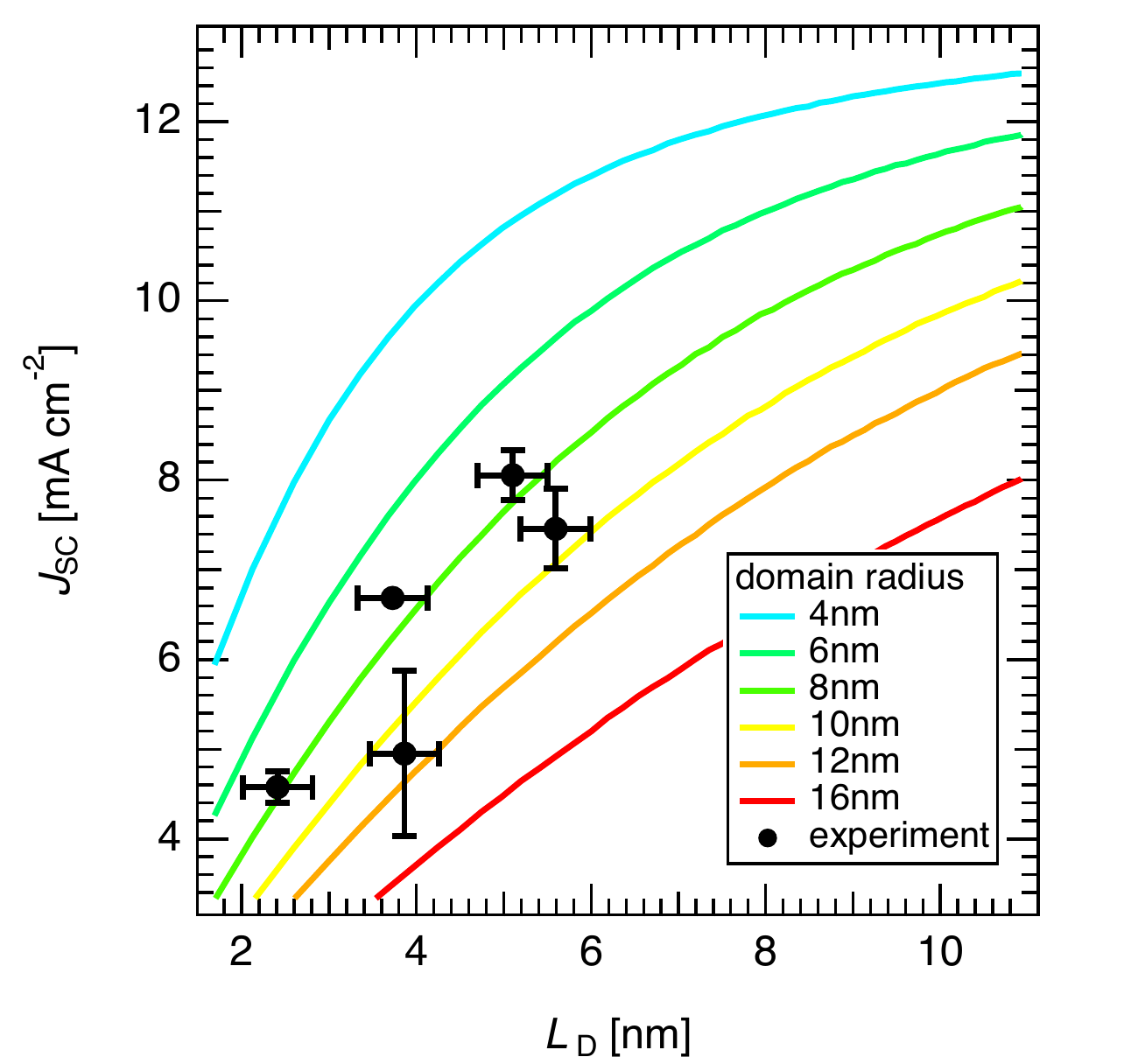}
\caption{Simulated dependence of short circuit current on exciton diffusion length for different sizes of a spherical polymer domain (solid lines) compared to experimental data (symbols).}
\label{IQE}
\end{figure}

To verify the connection between limited exciton diffusion and reduced $J_{\mathrm{sc}}$ in organic solar cells, we measured bulk heterojunction PCDTBT:PC$_{71}$BM solar cells prepared from the same PCDTBT batches used for studying exciton diffusion. At high molecular weights ($M_{\mathrm{n,SEC}} > 27.6\,\mathrm{kg\,mol^{-1}}$), we found it difficult to prepare solar cells due to high viscosity of the polymer solution. We compared EQE and  light transmission measurements of all devices to identify deviating active layers. Batch 5 ($50.7\,\mathrm{kg\,mol^{-1}}$) could not be fully dissolved and was therefore not prepared. Figure \ref{ldens} shows  $J_{\mathrm{sc}}$ of solar cells with comparable active layer thicknesses. The  trend is similar to  $L_{\mathrm{D}}$. To confirm the relation between  $J_{\mathrm{sc}}$ and $C_{\mathrm{cr}}$ of our limited set of data, we applied our model to data of different PCDTBT batches studied by Lombeck \textit{et al.}\cite{Lombeck2016} We indeed found a similar linear trend comparable to our results (see Supporting Information). To examine a possible correlation between diffusion length and short circuit current, we simulated the process of charge generation within a simplified bulk heterojunction solar cell. In our model, we simulated the probability of excitons reaching the boundary of a spherical polymer domain. We performed a kinetic Monte Carlo simulation similar to Mikhnenko \textit{et al.}.\cite{Mikhnenko2012} Excitons were randomly generated within a spherically confined grid. After generation, each exciton performed a random walk with lifetime $\tau'$ that was generated from the mono-exponential distribution ($\tau_0 = 750$\,ps).\cite{Mikhnenko2012,Anthanasopoulos2009}  Diffusion lengths were varied by changing the diffusion coefficient between $[0.05-5.0]\,\cdot\mathrm{10^{-4}cm^{2}s^{-1}}$ according to equation (\ref{length}). The ratio $Q$  of excitons reaching the domain boundary within $\tau'$ and generated excitons was determined as a function of $L_{\mathrm{D}}$.  We modelled a relative $J_{\mathrm{sc}}$  by assuming perfect exciton dissociation and charge extraction. $J_{\mathrm{sc}}$ then is proportional to $Q$ from the simulation. We scaled our maximal value of $Q=1$ to $J_{\mathrm{sc}} = 13\,\mathrm{mAcm^{-2}}$ representing a typical experimental value\cite{Lombeck2016} of an optimized PCDTBT:PC$_{71}$BM device. Figure \ref{IQE} shows the simulated connection between $J_{\mathrm{sc}}$ and $L_{\mathrm{D}}$ for different domain sizes compared to our experimental results. In the region where  $L_{\mathrm{D}}$ is limited to only few nanometers, $J_{\mathrm{sc}}$ is proportionally dependent on $L_{\mathrm{D}}$, thus strongly limited. The simulation additionally shows that diffusion length becomes less important when it reaches values in the range of the domain size. Since we found no such saturation in the experimental data, it is likely that there is no further potential for increased $J_{\mathrm{sc}}$ with larger domain sizes in PCDTBT. Terao \textit{et al.} found a linear correlation between $L_{\mathrm{D}}$ and $J_{\mathrm{sc}}$ in small molecule--fullerene solar cells.\cite{terao2007} Our experimental and simulated results indicate that this correlation also exists in polymer-based solar cells for exciton diffusion lengths smaller than the domain size.

\section{Conclusion}

We found carbazole homocoupling defects and short polymer chain lengths to decrease the singlet exciton diffusion length in PCDTBT. Concentrations of 10\,\%  homocoupling decreased the  diffusion coefficient by one order of magnitude. Short polymer chains below $M_{\mathrm{n}}=10\,\mathrm{kg\,mol^{-1}}$  impaired exciton diffusion in films with 0\,\% and 6\,\% homocoupling. We concluded that homocoupling defects divide the polymer chain into separate chromophores. Size and distribution of these chromophores depend on homocoupling concentration and molecular weight of the polymer. We simulated the average chromophore length and density according to our experimental polymer parameters (hc, $M_{\mathrm{n,SEC}}$).  Exciton diffusion length was significantly lower in batches with average conjugation lengths below 10 CDTBT repeat units. Such short chromophores occur in homocoupling defect-free films with low molecular weight below $10\,\mathrm{kg\,mol^{-1}}$ as well as in  films with high homocoupling concentration. The exciton diffusion length decreased linearly with increasing chromophore density in the film. We found a similar correlation for $J_{\mathrm{sc}}$ of solar cells prepared from the same PCDTBT batches. Our kinetic Monte Carlo simulation supported the idea that decreased $J_{\mathrm{sc}}$ is caused by limited exciton diffusion, explaining findings from the literature.\cite{Hendriks2014,Lombeck2016,Kingsley2014,Wakim2009}

\section{Methods}

The PCDTBT batches were synthesized as described in the Supporting Information. Carbazole homocoupling defects were intentionally introduced with varying amounts of an asymmetric carbazole monomer, similar to the synthesis procedure described previously.\cite{Lombeck2016}  The homocoupling concentration was determined by high-temperature NMR measurements. Molecular weights have been determined by high-temperature size exclusion chromatography in trichlorobenzene at $150\,^{\circ}\mathrm{C}$. Molecular weight, dispersity, homocoupling concentration and experimental values are given in the Supporting Information. Conventional PCDTBT with 2.5\,\% hc was purchased from 1-material and used without further purification. All batches were dissolved in 1,2-dichlorobenzene (DCB) at 5\,mg/ml and stirred for 3 days in nitrogen atmosphere. PCBM dissolved in DCB was added to the polymer solutions in different ratios. Films of 800\,nm thickness were produced by drop-coating $10\,\mu$l of solution on 5$\times5\,\mathrm{mm}^2$ silicon substrates with 100\,nm  thermally grown $\mathrm{SiO}_2$. Thick films were chosen to minimize the influence of surface quenching. As a test of our experiment, we additionally measured exciton diffusion in spin-coated films with 50\,nm thickness. We found comparable values for $D$. Values for $\tau_0$ were comparable between 50-800\,nm but significantly decreased when films were thinner than 50\,nm due to surface quenching (see Supporting Information).\cite{Lin2014} PL decay was acquired with a home-built confocal laser scan microscope. Samples were excited with a pulsed and filtered supercontinuum light source (SuperK EXR-15 with SELECTplus filter from  NKT Photonics) at $\lambda = 532\, $nm with a repetition rate of 19.5\,MHz at an pulse energy density of $0.5\,\mathrm{\mu J\,cm^{-2}}$. Two films of each polymer and PCBM concentration were measured at 10 different $100\times100\,\mu\mathrm{m}^2$ regions by scanning over the sample with a Zeiss LWD $63\times,N\!A=0.75$ objective,  collecting the PL with a single photon counting avalanche photodiode  (Micro Photon Devices, MPD-050-CTB). The instrument response function of the system is 53\,ps (full width at half maximum). PL lifetimes were consistent in all regions of the samples and between different films of the same polymer batch. Solar cells were prepared by separately dissolving all PCDTBT batches ($12\,\mathrm{mg\,ml^{-1}}$) and PC$_{71}$BM ($48\,\mathrm{mg\,ml^{-1}}$) in DCB. PCDTBT and PC$_{71}$BM solutions were mixed with a weight ratio of 1:4. Glass/ITO substrates were cleaned in Hellmanex\textregistered, aceton, isopropanol, deionized water and dried under nitrogen stream. $\textrm{{PEDOT:PSS}}$ was spincoated (30\,nm) onto the substrates and heated at $130^{\circ}\,\mathrm{C}$ for 20\,min. The blend solution was spincoated in nitrogen atmosphere. Calcium (2\,nm) and aluminum (150\,nm) were evaporated on the film as contacts. Short circuit currents under illumination of 1\,sun were calculated from calibrated external quantum efficiency measurements (EQE). Absolute EQE spectra were obtained by illuminating solar cells with a chopped tunable light source measuring the photocurrent with a lock-in amplifier (Stanford Research Systems SR830). A calibrated silicon photodiode was measured simultaneously as reference.

\begin{acknowledgements}
The authors are grateful to H. Komber, IPF Dresden, for determining hc contents, and to J. Brandt and A. Lederer, IPF Dresden, for providing HT-SEC measurements.  This work was partly funded by the DFG (project DE830/17-1).
\end{acknowledgements}

\appendix

\section{Corresponding Author}
Email: deibel@physik.tu-chemnitz.de

\section{Author contributions}
M. Streiter measured exciton diffusion, simulated effective conjugation lengths, interpreted the results and wrote the manuscript. D. Beer performed Monte Carlo simulations  and prepared samples. F. Meier assisted with measurements. C. Lienert and F. Lombeck synthesized PCDTBT of varying homocoupling concentration. M. Sommer supervised the synthetic work and was involved in discussions and writing the manuscript. C. Deibel supervised the experimental work, simulations and contributed to interpretation of results and writing of the manuscript.

\section{Notes}
The authors declare no competing financial interest.

\section{Supplementary Information}
Synthesis of PCDTBT, analysis of molecular weight and homocoupling, PL decay, Stern-Volmer plots, derivation of equations, comparison to literature values

\end{document}


\section{PCDTBT data}

\begin{table}
    \centering
\caption{List of  measured PCDTBT batches  with  different   molecular weight and  carbazole homocoupling (hc) concentration. }

\begin{tabular}{cccccccc} 
batch   & lab ID &$M_{\mathrm{n,SEC}}$  &hc                 & $\tau_0$      & $D$ & $N_{\mathrm{eff}}$ (sim.)&$M_{\mathrm{eff}}$ (calc.)\\ 
   no.  &        &$[\mathrm{kg\,mol^{-1}}]$              &$[\mathrm{mol}\%]$ &  [ps] &$[10^{-4}\mathrm{cm^2 s^{-1}}]$  & [CDTBT]   &$[\mathrm{kg\,mol^{-1}}]$\\
\hline
1   &CL10\_CF           &6.2    & 0     & 724   & 0.32  &  8.8 & 6.2\\ 
2   &CL26\_CF           &9.4    & 6     & 644   & 0.15  &  5.3 & 3.6\\ 
3   &FL498              &31.0   & 10    & 699   & 0.06  &  4.3 &3.1\\ 
4   &CL24               &27.6   & 0     & 828   & 0.63  &  39.3 &27.6\\ 
5   &CL12               &50.7   & 0     & 862   & 0.50  &  72.2 &50.7\\
6   &Commercial         &22.3   & 2.5   & 850   & 0.51  &  12.1 &8.6\\ 
7   &CL27               &45.8   & 6     & 672   & 0.37  &  7.2 &5.2\\

\end{tabular}
\label{sample_hc}
\end{table}

\section{Instruments}
\subsection*{High temperature size exclusion chromatography (HT-SEC)}
\begin{figure}[h]
  \includegraphics*[scale=0.6]{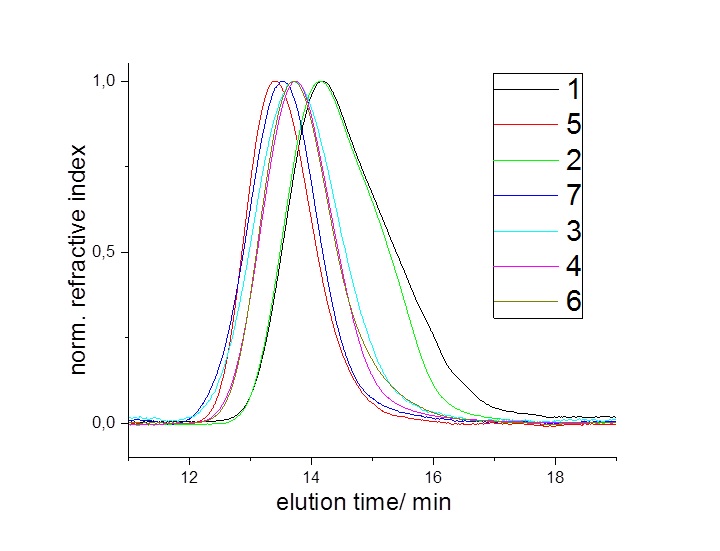}
    \caption{HT-SEC of samples 1-7 in trichlorobenzene at $150\,^{\circ}\mathrm{C}$.}
    \label{sec}
    \end{figure}

HT-SEC was performed on an Agilent PL GPC 220 in 1,2,4 trichlorobenzene at $150\,^{\circ}\mathrm{C}$ and a flow rate of 1\,ml/min. Two 2x Agilent Mixed-B-LS columns were used for separation. Polystyrene calibration was used. HT-SEC spectra are shown in figure (\ref{sec}).

\subsection*{NMR spectroscopy}
\begin{figure}[h]
  \includegraphics*[scale=0.9]{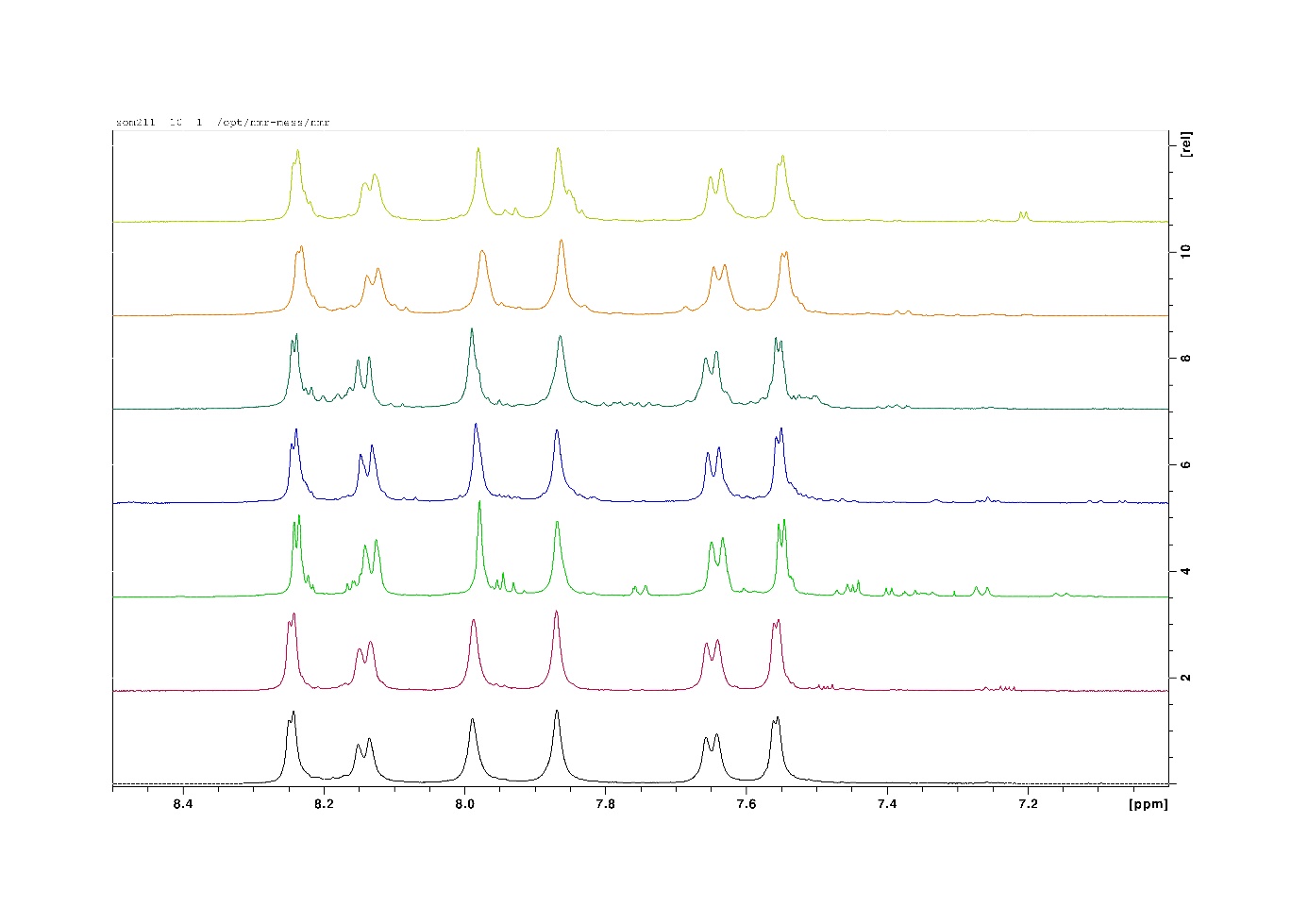}
    \caption{500\,MHz $^1$H-NMR ($\mathrm{C_2 D_2 Cl_4}$, $120\,^{\circ}\mathrm{C}$) of samples 1-7 (from top to bottom) showing the aromatic region. The signal at $\approx8.2$\,ppm arises from carbazole homocoupling and was used for hc content determination.  }
    \label{nmr}
    \end{figure}

$^1$H NMR spectroscopy (500.13\,MHz) of polymers was performed with a Bruker Avance III spectrometer using a 5\,mm BBI gradient probe at $120\,^{\circ}\mathrm{C}$. $\mathrm{C_2 D_2 Cl_4}$ ($\delta(^1\mathrm{H}) = 5.98\,\mathrm{ppm}$)  was used as solvent. NMR spectra are shown in figure (\ref{nmr}).

\subsection*{Photoluminescence spectroscopy}
Steady state photoluminescence spectra were measured in dichlorobenzene solution (polymer concentration  $0.025\,\mathrm{mg/ml}$) with a Varian Cary Eclipse spectrometer in  quartz cuvettes. The excitation wavelength was 532\,nm. 

\section*{Synthesis of PCDTBT}

\subsection*{Batch 1}

4,7-Bis(5-bromothiophen-2-yl)benzo[c][1,2,5]thiadiazole (TBT-Br$_2$) (627\,mg, 1.37\,mmol) and 9-(heptadecan-9-yl)-2,7-bis(4,4,5,5-tetramethyl-1,3,2-dioxaborolan-2-yl)-9H-carbazole  (Cbz-Bpin$_2$) (900\,mg, 1.37\,mmol) were loaded in a 100\,ml screw cap vial. Under argon atmosphere the catalyst Pd(PPh$_3$)$_4$ was added and the vial closed with a new septum. Toluene (27.4\,ml) and an aqueous 2M K$_2$CO$_3$ solution (27.4\,ml) were purged with nitrogen for 20\,min. The solvent mixture and 3\,drops Aliquat 336\textregistered \,were added to the reaction vial. The septum was displaced by a screw cap under a stream of nitrogen and the reaction mixture was heated at $80\,^{\circ}\mathrm{C}$ under vigorous stirring. After 112\,h, the mixture was allowed to cool to room temperature. The aqueous layer was removed and 1,2-dichlorobenzene was added to the organic layer to dissolve solid components. The resulting polymer was precipitated into methanol, filtered and washed by Soxhlet extraction with methanol, acetone, hexanes, chloroform, and chlorobenzene. The chloroform fraction was collected, filtered over silica gel and concentrated under vacuum. The polymer was precipitated into methanol and dried under vacuum.

\noindent  Yield (chloroform fraction): 87\,\%\\
(SEC, TCB, $150\,^{\circ}\mathrm{C}$) $M_{\mathrm{n}}=$ 6.2\,kg/mol, $M_{\mathrm{w}}/M_{\mathrm{n}}$ = 5.32, hc: 0\,mol\%

\subsection*{Batch 2}

TBT-Br$_2$ (589\,mg, 1.23\,mmol), Cbz-Bpin$_2$ (846\,mg, 1.23\,mmol), and 2-bromo-9-(heptadecan-9-yl)-7-(4,4,5,5-tetramethyl-1,3,2-dioxaborolan-2-yl)-9H-carbazole (Bpin-Cbz-Br)  \\  (100\,mg, 0.164\,mmol) were loaded in a 100\,ml screw cap vial. Under argon atmosphere the catalyst Pd(PPh$_3$)$_4$ was added and the vial closed with a new septum. Toluene (27.4\,ml) and an aqueous 2M K$_2$CO$_3$ solution (27.4\,ml) were purged with nitrogen for 20\,min. The solvent mixture and 3\,drops Aliquat 336\textregistered \,were added to the reaction vial. The septum was displaced by a screw cap under a stream of nitrogen and the reaction mixture was heated at  $80\,^{\circ}\mathrm{C}$ under vigorous stirring. After 24\,h, the mixture was allowed to cool to room temperature. The aqueous layer was removed and 1,2-dichlorobenzene was added to the organic layer to dissolve solid components. The resulting polymer was precipitated into methanol, filtered and washed by Soxhlet extraction with methanol, acetone, hexanes, chloroform, and chlorobenzene. The chloroform fraction was collected, filtered over silica gel and concentrated under vacuum. The polymer was precipitated into methanol and dried under vacuum. 

\noindent Yield (chloroform fraction): 89\,\%\\
(SEC, TCB, $150\,^{\circ}\mathrm{C}$) $M_{\mathrm{n}}=$ 9.4\,kg/mol, $M_{\mathrm{w}}/M_{\mathrm{n}}$ = 3.86, hc: 6\,mol\%

\subsection*{Batch 3} 

TBT-Br$_2$ (300\,mg, 0.65\,mmol), Cbz-Bpin$_2$ (431\,mg, 0.65\,mmol), and BpinCbzBr (88.9\,mg, 0.14\,mmol) were loaded in a 100\,ml screw cap vial. Under argon atmosphere the catalyst Pd(PPh$_3$)$_4$ was added and the vial closed with a new septum. Toluene (27.4\,ml) and an aqueous 2M K$_2$CO$_3$ solution (27.4\,ml) were purged with nitrogen for 20\,min. The solvent mixture and 3\,drops Aliquat 336\textregistered \, were added to the reaction vial. The septum was displaced by a screw cap under a stream of nitrogen and the reaction mixture was heated at  $80\,^{\circ}\mathrm{C}$ under vigorous stirring. After the reaction was completed, the mixture was allowed to cool to room temperature. The aqueous layer was removed and 1,2-dichlorobenzene was added to the organic layer to dissolve solid components. The resulting polymer was precipitated into methanol, filtered and washed by Soxhlet extraction with methanol, acetone, hexanes, chloroform, and chlorobenzene. The chlorobenzene fraction was collected and dried under vacuum.

\noindent  Yield (chloroform fraction): 86\,\%\\
(SEC, TCB, $150\,^{\circ}\mathrm{C}$) $M_{\mathrm{n}}=$ 31.0\,kg/mol, $M_{\mathrm{w}}/M_{\mathrm{n}}$ = 2.90, hc: 10\,mol\%

\subsection*{Batch 4} 
TBT-Br$_2$ (627\,mg, 1.37\,mmol) and Cbz-Bpin$_2$ (900\,mg, 1.37\,mmol) were loaded in a 100\,ml screw cap vial. Under argon atmosphere the catalyst Pd(PPh$_3$)$_4$ was added and the vial closed with a new septum. Toluene (27.4\,ml) and an aqueous 2M K$_2$CO$_3$ solution (27.4\,ml) were purged with nitrogen for 20\,min. The solvent mixture and 3\,drops Aliquat 336\textregistered \, were added to the reaction vial. The rubber seal was replaced by a Teflon seal under a stream of nitrogen and the reaction mixture was heated at  $80\,^{\circ}\mathrm{C}$ under vigorous stirring. After 72\,h the mixture was allowed to cool to room temperature. The aqueous layer was removed and the resulting polymer was precipitated into methanol, filtered and washed by Soxhlet extraction with methanol, acetone, hexanes, chloroform, and chlorobenzene. The polymer was precipitated into methanol, filtered and dried under vacuum.

\noindent  Yield (chloroform fraction): 57\,\%\\
(SEC, TCB, $150\,^{\circ}\mathrm{C}$) $M_{\mathrm{n}}=$ 27.6\,kg/mol, $M_{\mathrm{w}}/M_{\mathrm{n}}$ = 2.86, hc: 0\,mol\%

\subsection*{Batch 5}  

TBT-Br$_2$ (100\,mg, 0.218\,mmol), Cbz-Bpin$_2$ (143\,mg, 0.218\,mmol), the catalyst precursor Pd$_2$dba$_3$ (1.0\,mg, 0.5\,mol\,\%), and the ligand PPh$_3$ (11.4\,mg, 20\,mol\,\%) were loaded in an 18\,ml screw cap vial. Toluene (8.7\,ml) and an aqueous 2M K$_2$CO$_3$ solution (8.7\,ml) were purged with nitrogen for 20\,min. The solvent mixture and 1 drop Aliquat 336\textregistered \, were added to the reaction vial under nitrogen atmosphere. The septum was displaced by a screw cap under a stream of nitrogen and the reaction mixture was heated at $80\,^{\circ}\mathrm{C}$ under vigorous stirring. After 72\,h, the mixture was allowed to cool to room temperature. The aqueous layer was removed. The resulting polymer in the organic layer was precipitated into methanol, filtered and washed by Soxhlet extraction with methanol, acetone, hexanes, chloroform, and chlorobenzene. The chlorobenzene fraction was poured into methanol to precipitate the polymer. The solvent was decanted and the polymer was dried under vacuum.

\noindent  Yield (chloroform fraction): 87\,\%\\
(SEC, TCB, $150\,^{\circ}\mathrm{C}$) $M_{\mathrm{n}}=$ 50.7\,kg/mol, $M_{\mathrm{w}}/M_{\mathrm{n}}$ = 2.25, hc: 0\,mol\%

\subsection*{Batch 7}  

TBT-Br$_2$ (589\,mg, 1.23\,mmol), Cbz-Bpin$_2$ (846\,mg, 1.23\,mmol), and BpinCbzBr (100\,mg, 0.164\,mmol) were loaded in a 100\,ml screw cap vial. Under argon atmosphere the catalyst Pd(PPh$_3$)$_4$ was added and the vial closed with a new septum. Toluene (27.4\,ml) and an aqueous 2M K$_2$CO$_3$ solution (27.4\,ml) were purged with nitrogen for 20\,min. The solvent mixture and 3\,drops Aliquat 336\textregistered \, were added to the reaction vial. The septum was displaced by a screw cap under a stream of nitrogen and the reaction mixture was heated at $80\,^{\circ}\mathrm{C}$ under vigorous stirring. After the reaction was completed, the mixture was allowed to cool to room temperature. The aqueous layer was removed and 1,2-dichlorobenzene was added to the organic layer to dissolve solid components. The resulting polymer was precipitated into methanol, filtered and washed by Soxhlet extraction with methanol, acetone, hexanes, chloroform, and chlorobenzene. The chlorobenzene fraction was precipitated into methanol, the mixture centrifuged and the solids dried under vacuum.

\noindent Yield (chloroform fraction): 53\,\%\\
(SEC, TCB, $150\,^{\circ}\mathrm{C}$) $M_{\mathrm{n}}=$ 45.8\,kg/mol, $M_{\mathrm{w}}/M_{\mathrm{n}}$ = 2.58, hc: 6\,mol\%

\section{Derivation of effective conjugation length}

The effective conjugation length $N_{\mathrm{eff}}$ and effective molecular weight $M_{\mathrm{eff}}$ can be estimated by assuming homocoupling defects to be spaced equidistant within the polymer chain. In our model, the chromophore consists of AB segments unbroken by homocoupled ABB segments.\\
Example: A-B-A-B-\textbf{A-B-B}-A-B-A-B-\textbf{A-B-B}-A-B-A-B\\

\noindent Homocoupling concentration hc is defined as the number of homocoupling bonds (2 in the example) divided by the total number of bonds between all units (17 in the example), $\textrm{hc} = 2/17 \approx 12\,\%$. In the example, the chain with $N_0 = 8$ units (6$\times$AB, 2$\times$ABB) consists of three chromophores with $N=2 \Rightarrow N_{\mathrm{eff}} = 2$. The amount of ABB units $N_{\mathrm{hc}}$ can be estimated  with $N_{\mathrm{hc}} \approx 2\textrm{hc}N_0$. The chain is divided $N_{\mathrm{hc}}$ times and the segments available for building up chromophores is limited to $N'_0 = N_0 - N_{\mathrm{hc}}$. In the example, 6 AB units are divided by 2 ABB units into 3 chromophores. This can be expressed as:

\begin{equation}
    N_{\mathrm{eff}} = \frac{N_0 - 2\textrm{hc} \cdot N_0}{2\textrm{hc}N_0 +1} \approx \frac{N_0}{2\textrm{hc}N_0 + 1}
     \label{lengthN}
\end{equation}

\begin{figure}[h]
\includegraphics*[scale=0.7]{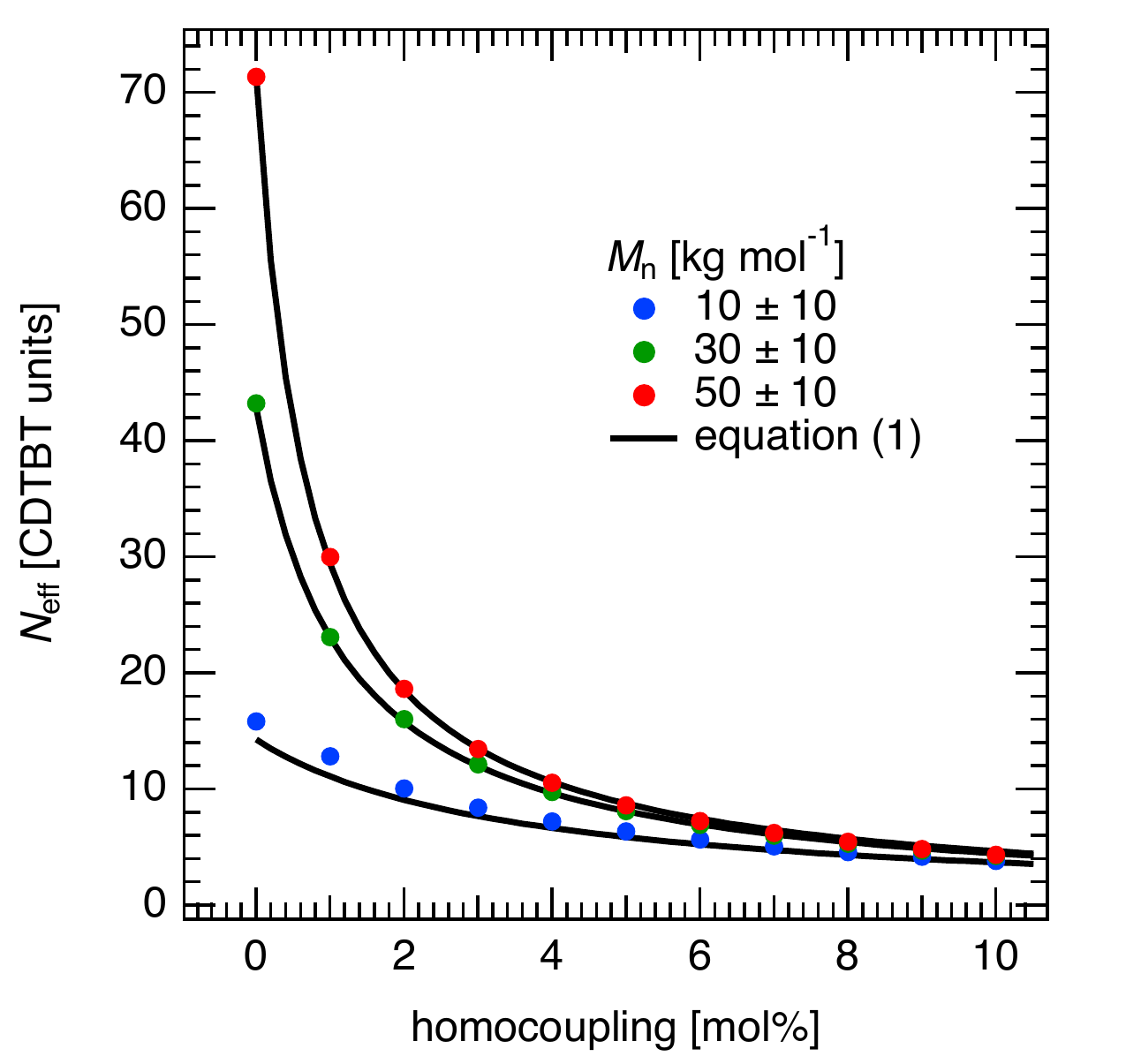}
\caption{The effect of homocoupling on the effective conjugation length for different number average molecular weights. Dots are from our simulation, black lines from equation (\ref{lengthN}). At low molecular weights, the analytical equation underestimates $N$.}
\label{sim_analyt}
\end{figure}

Note that this equation is unprecise for short chains (as in the example where $N_{\mathrm{hc}} = 1.888$ instead of 2). We improved the results for short chains (low molecular weights) by simulating $N_{\mathrm{eff}}$ as described in the paper. The equation   underestimates $N_{\mathrm{eff}}$ at a low molecular weights below $10\,\mathrm{kg\,mol^{-1}}$ compared to the simulation as shown in figure \ref{sim_analyt}. An effective mass of a homocoupled copolymer (corresponding to the  effective conjugation length) can roughly be estimated with $N\cdot M_{\mathrm{unit}}  \approx M_{\mathrm{n}}$. Here, $M_{\mathrm{unit}}$ is the molecular weight of the monomer A-B unit. Note that the changed molecular weight caused by additional B groups is neglected here. This is corrected in our simulation.

\begin{equation}
    M_{\mathrm{eff}}(\textrm{hc})\approx M_{\mathrm{n}}\left[\frac{2\textrm{hc}\cdot M_{\mathrm{n}}}{M_{\mathrm{unit}}}+1\right]^{-1}
     \label{lengthm}
\end{equation}

\noindent  The number of chromophores per volume in the film  $C_{\mathrm{cr}}$ can be expressed as
\begin{equation}
    C_{\mathrm{cr}} \approx \frac{\rho_{\mathrm{PCDTBT}}}{N_{\mathrm{eff}} \cdot M_{\mathrm{unit}}}.
\end{equation}

\begin{figure}[h]
\includegraphics*[scale=0.7]{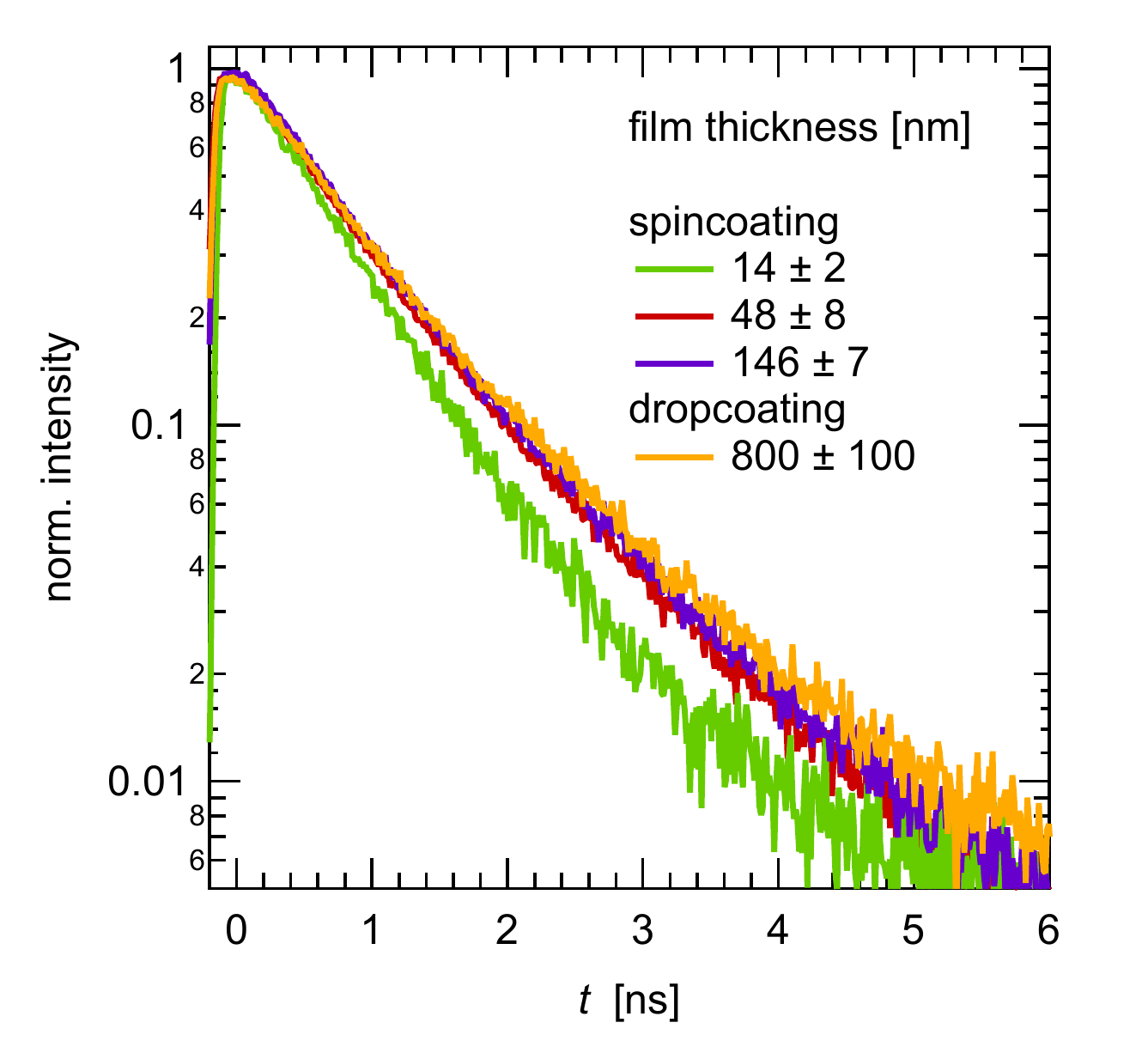}
\caption{Thickness dependent photoluminescence decay of a commerical PCDTBT batch. Below 50\,nm, surface quenching decreases the PL lifetime.}
\label{thickness}
\end{figure}

\begin{figure}[h]
\includegraphics*[scale=0.7]{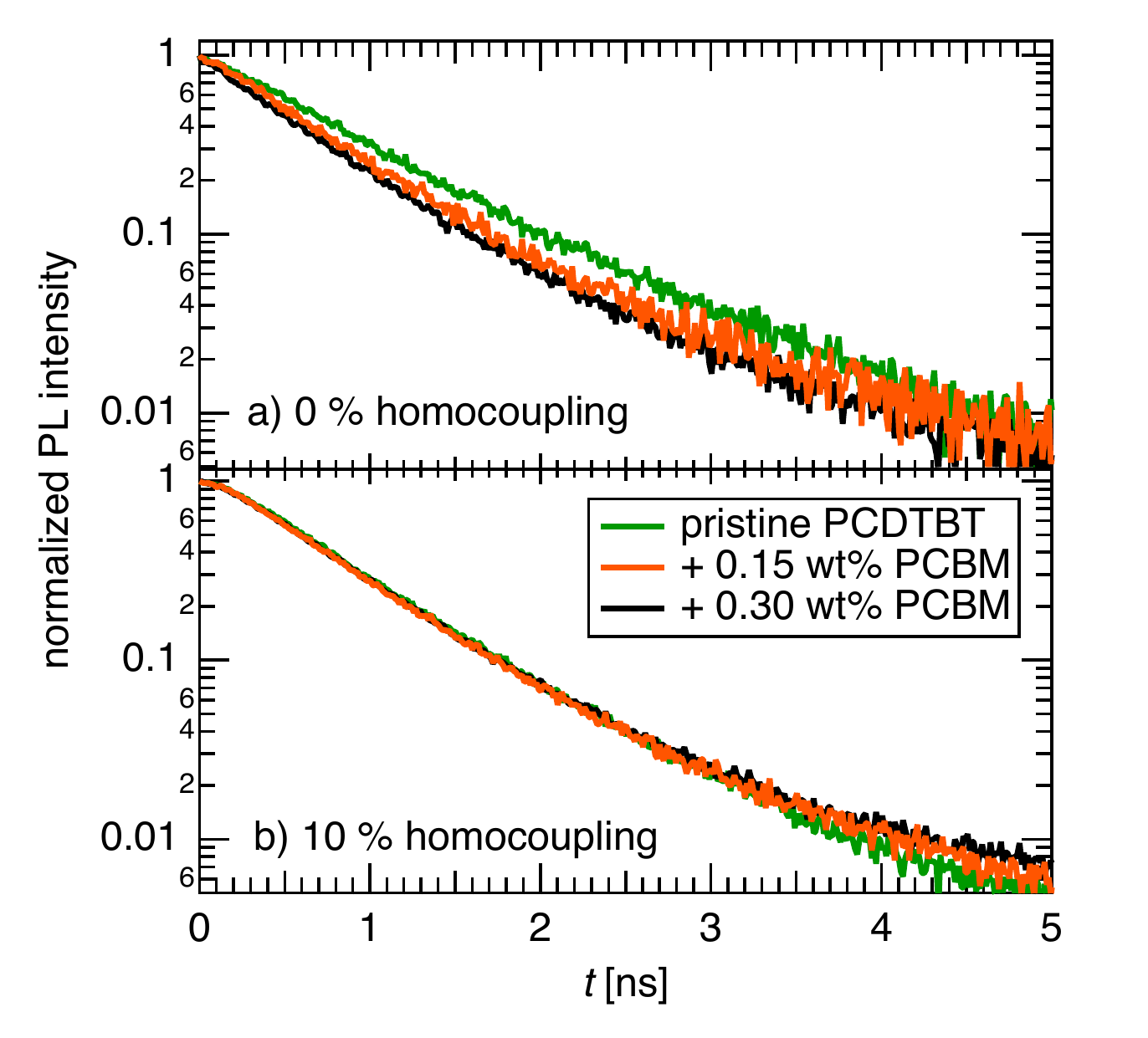}
\caption{PL decay of  PCDTBT films with $0\,\%$ and $10\,\%$ homocoupling. The lifetime becomes shorter with higher amounts of PCBM blended in the film at $0\,\%$ homocoupling. At $10\,\%$ homocoupling, the decay is comparable in the region between 0--4\,ns, indicating shorter exciton diffusion length.}
\label{decay}
\end{figure}

\begin{figure}[h]
\includegraphics*[scale=0.7]{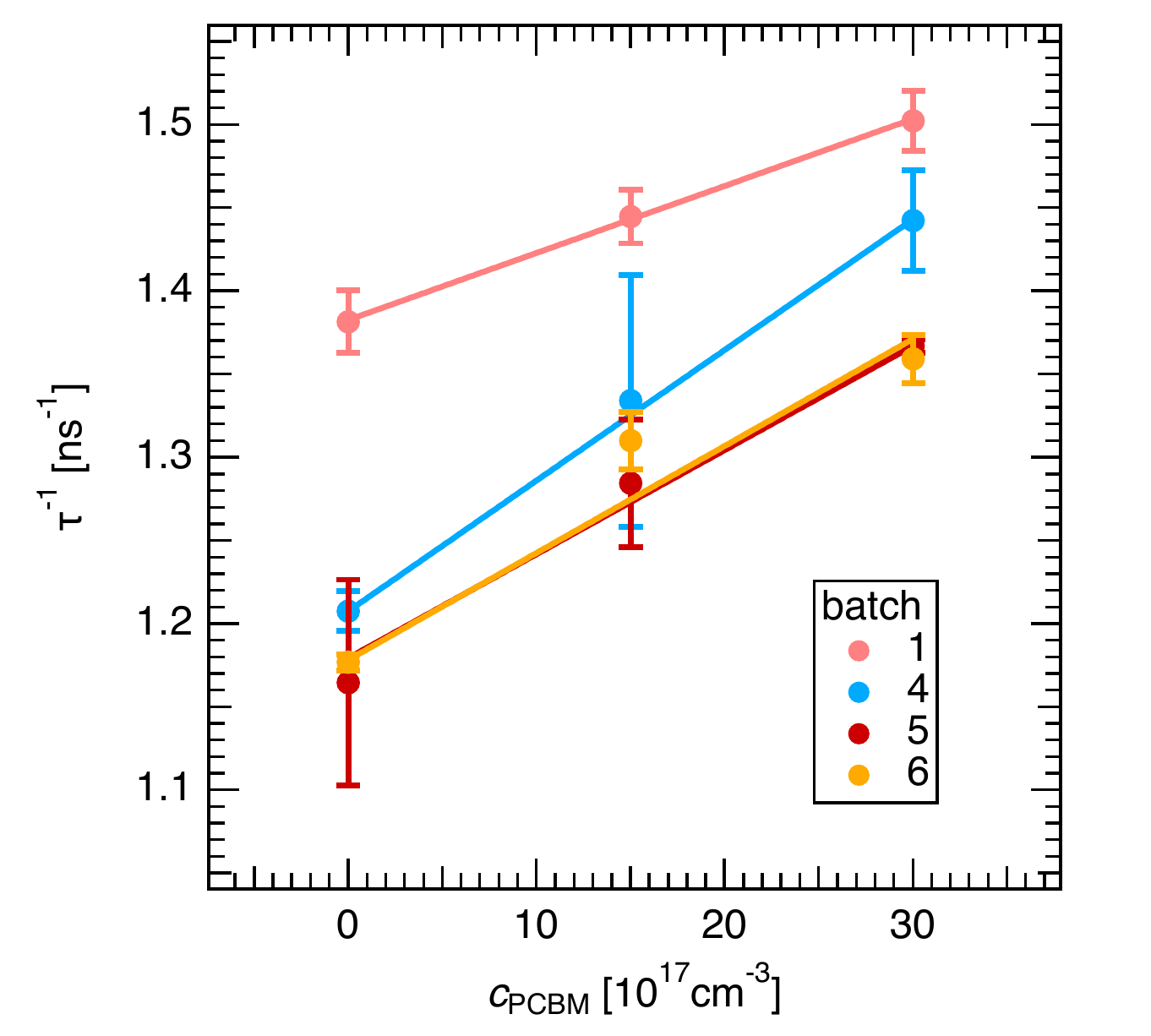}
\caption{Stern--Volmer plot of inverse PL lifetimes $\tau^{-1}$ over  quencher concentration $c$ in  batches of PCDTBT with low homocoupling concentrations.}
\label{sv_lo}
\end{figure}
 
\begin{figure}[h]
\includegraphics*[scale=0.7]{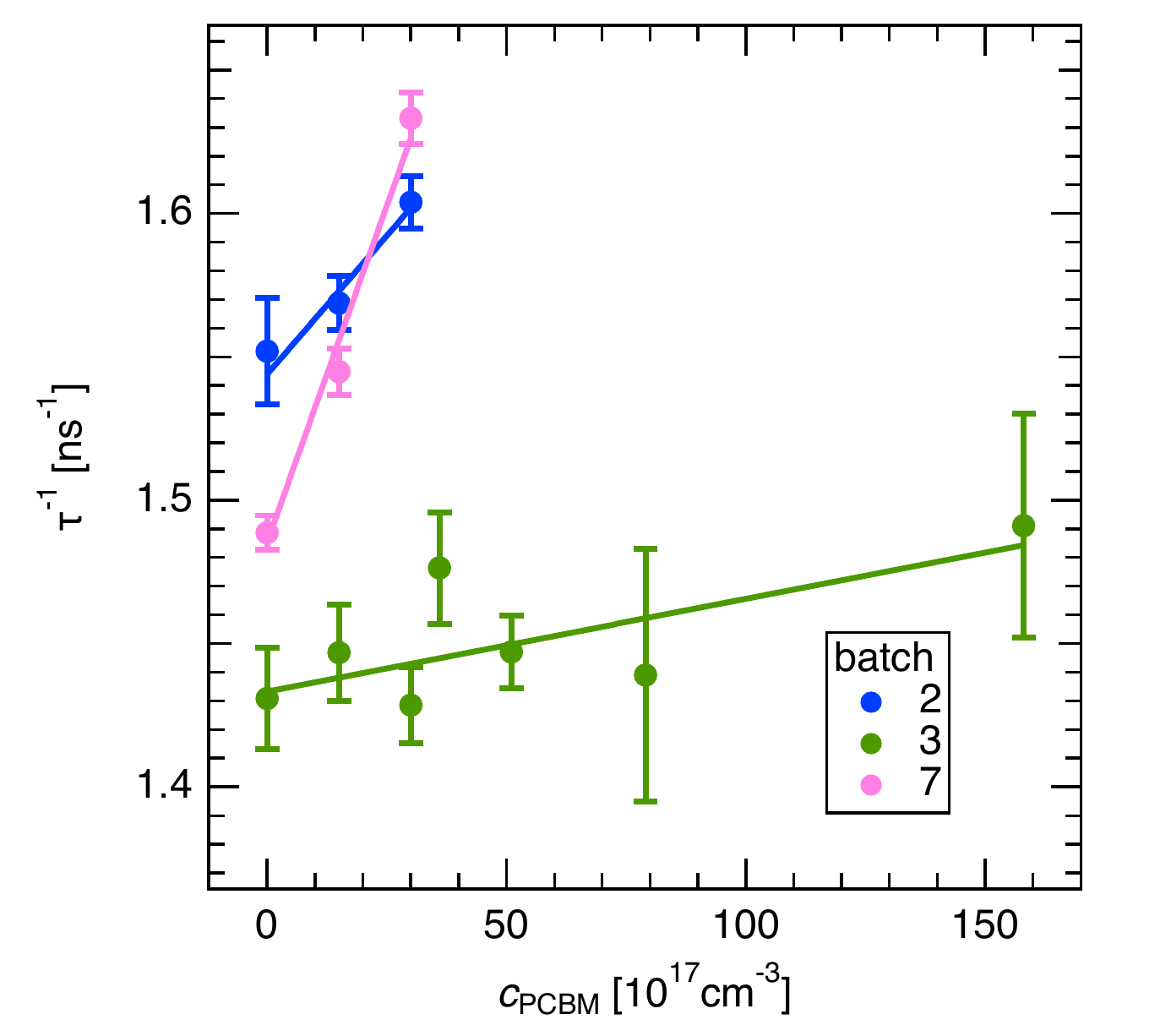}
\caption{Stern--Volmer plot of inverse PL lifetimes $\tau^{-1}$ over  quencher concentration $c$ in  batches of PCDTBT with high homocoupling concentrations.}
\label{sv_hi}
\end{figure}

\begin{figure}[h]
\includegraphics*[scale=0.7]{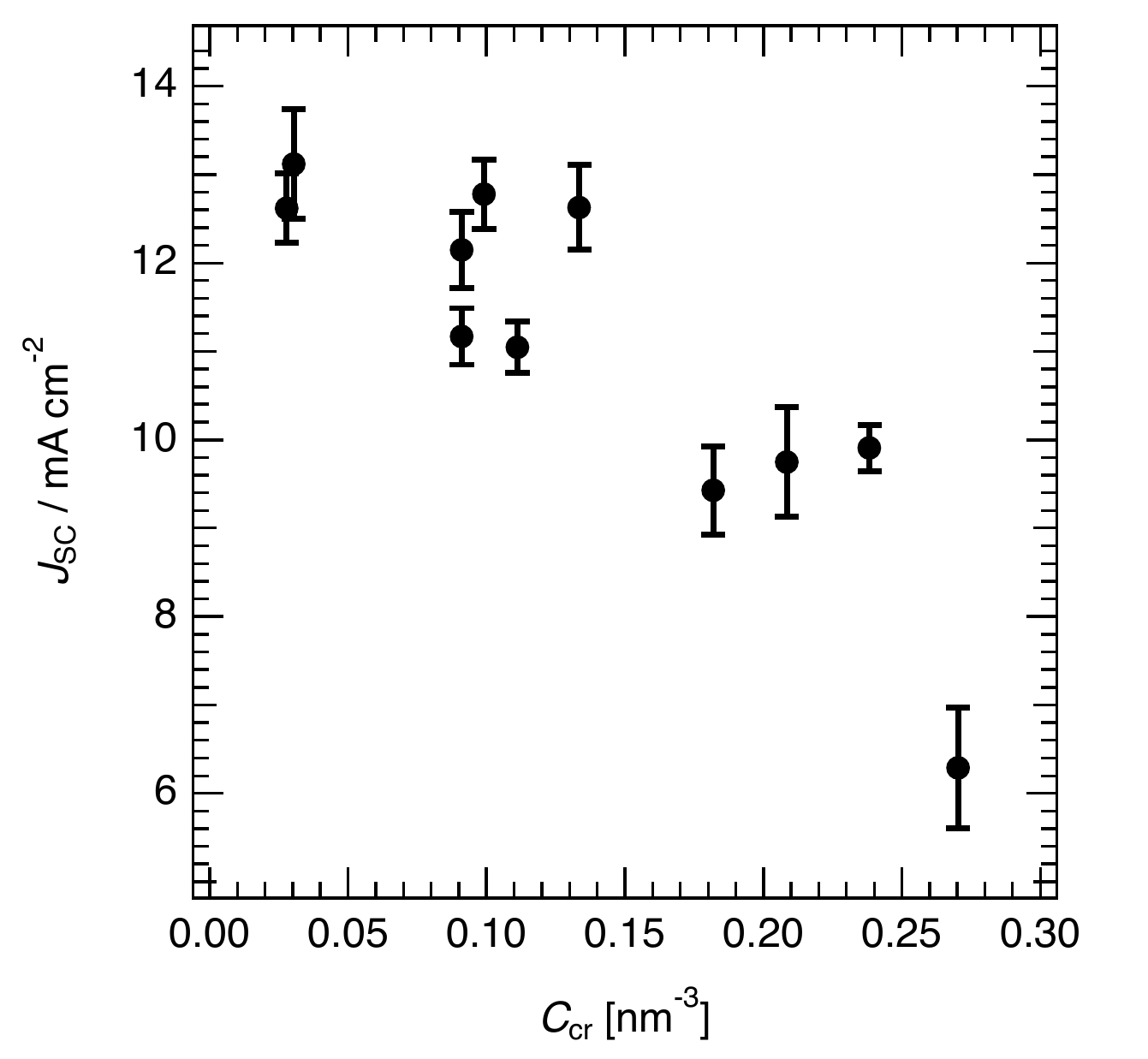}
\caption{Short circuit current of PCDTBT:PCBM solar cells with different homocoupling concentrations and molecular weights. The chromophore density $C_{\mathrm{cr}}$ was simulated as described in the paper. Data taken from Lombeck \textit{et al}., Adv. Energy Mater. \textbf{6}, 1 (2016).}
\label{lombeckcells}
\end{figure}